\documentclass[twocolumn,english,showpacs,superscriptaddress,prc,aps]{revtex4-1}
\usepackage[T1]{fontenc}
\usepackage[latin9]{inputenc}
\usepackage{babel}
\usepackage{amsmath}
\usepackage{amsthm}
\usepackage{amssymb}
\usepackage{graphicx}
\usepackage[colorlinks,citecolor=blue,linkcolor=blue]{hyperref}
\usepackage{breakurl}
\begin{document}

\title{New feature of low $p_{T}$ charm quark hadronization in $pp$ collisions at $\sqrt{s}=7$ TeV}

\author{Jun Song }

\affiliation{Department of Physics, Jining University, Shandong 273155, China}

\author{Hai-hong Li}

\affiliation{Department of Physics, Jining University, Shandong 273155, China}

\author{Feng-lan Shao}
\email{shaofl@mail.sdu.edu.cn}
\address{School of Physics and Engineering, Qufu Normal University, Shandong 273165, China}

\begin{abstract}
Treating the light-flavor constituent quarks and antiquarks that can well describe the data of light-flavor hadrons in $pp$ collisions at $\sqrt{s}=7$ TeV as the underlying source of chromatically neutralizing the charm quarks of low transverse momenta ($p_{T}$), we show that the experimental data of $p_{T}$ spectra of single-charm hadrons $D^{0,+}$, $D^{*+}$ $D_{s}^{+}$, $\Lambda_{c}^{+}$ and $\Xi_{c}^{0}$ at mid-rapidity in the low $p_{T}$ range ($2\lesssim p_{T}\lesssim7$ GeV/$c$) in $pp$ collisions at $\sqrt{s}=7$ TeV can be well understood by the equal-velocity combination of perturbatively-created charm quarks and those light-flavor constituent quarks and antiquarks. This suggests a possible new scenario of low $p_{T}$ charm quark hadronization, in contrast to the traditional fragmentation mechanism, in $pp$ collisions at LHC energies. This is also another support for the exhibition of the effective constituent quark degrees of freedom for the small parton system created in $pp$ collisions at LHC energies. 
\end{abstract}

\maketitle

\section{Introduction}

The experimental study of the quark-gluon plasma (QGP), a new state
of matter of QCD, is mainly through the heavy-ion collisions which
can create the big thermal parton system with relatively long lifetime.
Relative to heavy ion collisions, proton-nucleus ($pA$) collisions
create the ``intermediate'' parton system and proton-proton ($pp$)
collisions create the ``small'' parton system. The deconfined medium
is usually assumed to be not created in $pA$ and $pp$ collisions,
at least up to RHIC energies. In particular, the data of $pp$ collisions,
in the context of heavy-ion physics, are usually served as the baseline
to study the effects and/or properties of cold and hot nuclear matter
in $pA$ and $AA$ collisions, respectively. 

Large Hadron Collider (LHC) pushes the center of mass energy per colliding
nucleon up to TeV level, which brings new properties even for the
small parton system created in $pp$ collisions. Recent measurements
in $pp$ collisions at LHC energies from CMS and ALICE collaborations
find several remarkable similarities with heavy-ion collisions. In
high-multiplicity events of $pp$ collisions, the phenomena such as
long range angular correlations \cite{Khachatryan:2010gv,Khachatryan:2015lva}
and collectivity \cite{Khachatryan:2016txc,Ortiz:2014iva}, strangeness
enhancement \cite{Adam:2015vsf,ALICE:2017jyt}, and the increased
baryon to meson ratio at low transverse momentum ($p_{T}$) \cite{Abelev:2013haa,Adam:2016dau,Acharya:2017kfy}
are observed. These phenomena were already observed in heavy-ion collisions
at RHIC and LHC energies and are usually regarded as the typical behaviors
related to the formation of QGP. Theoretical studies of these striking
observations focus on what happens on the small parton system created
in $pp$ collisions at LHC energies through different phenomenological/theoretical
methods such as the mini-QGP creation or phase transition \cite{Liu:2011dk,Werner:2010ss,Bzdak:2013zma,Bozek:2013uha,Prasad:2009bx,Avsar:2010rf},
multiple parton interaction \cite{Sjostrand:1987su}, string overlap
and color re-connection at hadronization \cite{Bautista:2015kwa,Bierlich:2014xba,Ortiz:2013yxa,Christiansen:2015yqa},
etc. 

In the recent work \cite{Gou:2017foe}, we found that the mid-rapidity
data of $p_{T}$ spectra of light-flavor hadrons in the low $p_{T}$
range ($p_{T}\lesssim6$ GeV/$c$) in $pp$ collisions at $\sqrt{s}=7$
TeV can be well understood by the equal-velocity combination of up/down
and strange quarks and antiquarks with constituent masses at hadronization.
This suggests that the constituent quark degrees of freedom (CQdof)
play an important role in low $p_{T}$ hadron production in $pp$
collisions at LHC energies, which indicates the possible existence
of the underlying source with soft CQdof, a kind of new property for
the small parton system created in $pp$ collisions at LHC energies. 

The hadronization of charm quarks in $pp$ collisions is usually described
by the traditional fragmentation mechanism or fragmentation function.
In this paper, we study the possibility of new feature for the hadronization
of low $p_{T}$ charm quarks in $pp$ collisions at LHC energies.
As the aforementioned discussion, the production of light-flavor hadrons
in $pp$ collisions at LHC energies can be well described by the combination
of light-flavor constituent quarks and antiquarks of low $p_{T}$.
These constituent quarks and antiquarks also serve as an underlying
source for the color neutralization of charm quarks at hadronization
to form the single-charm hadrons. Specifically, the charm quark can
pick up a co-moving light anti-quark or two co-moving quarks to form
a single-charm meson or baryon, where the momentum characteristic
is the combination $p_{H}=p_{c}+p_{\bar{q},qq}$. This (re-)combination
mechanism of charm quark hadronization will reflect in the momentum
spectra of charm hadrons and, in particular, the ratio of charm baryon
to charm meson. Therefore, in this paper, we study the mid-rapidity
$p_{T}$ spectra of single-charm mesons $D^{0,+}$, $D^{*+}$, $D_{s}^{+}$
and baryons $\Lambda_{c}^{+}$, $\Xi_{c}^{0}$ and the ratios among
them in the framework of quark (re-)combination mechanism (QCM), and
compare our results with available experimental data and several theoretical
predictions by fragmentation mechanism. 

The paper is organized as follows: Sec. II will introduce a working
model in quark (re-)combination mechanism for charm quark hadronization.
Sec. III presents our results and relevant discussions. Summary is
given at last in Sec. IV. 

\section{charm quark hadronization in QCM}

The (re-)combination mechanism of charm quark hadronization was proposed
in early 1980s \cite{Das:1977cp,Hwa:1979pn,Chiu:1978nc} and has many
applications in both hadron-hadron collisions \cite{Hwa:1994uha,Cuautle:1997ti,Braaten:2002yt}
and relativistic heavy ion collisions \cite{Greco:2003vf,Fries:2008hs,Cao:2013ita,Prino:2016cni}.
Because of the lack of the sufficient knowledge for the spatial information
of the small parton system created in $pp$ collisions at LHC energies,
in this section, we present a working model for the (re-)combination
hadronization of charm quarks in the low $p_{T}$ range in momentum
space, which only incorporates the most basic feature of QCM, i.e.,
the equal-velocity combination approximation. The unclear non-perturbative
dynamics such as the selection of different spin states and the formation
competition between baryon and meson in the combination are treated
as model parameters. 

\subsection{formulas in momentum space}

The momentum distributions of the single-charm meson $M_{c\bar{l}}$
and baryon $B_{cll'}$ in QCM, as formulated in e.g. \cite{Wang:2012cw,Li:2017zuj}
in general, can be obtained by 
\begin{align}
f_{M_{c\bar{l}}}(p) & =\int dp_{1}dp_{2}f_{c\bar{l}}(p_{1},p_{2})\,{\cal R}_{M_{c\bar{l}}}(p_{1},p_{2};p),\label{eq:fmc}\\
f_{B_{cll'}}(p) & =\int dp_{1}dp_{2}dp_{3}f_{cll'}(p_{1},p_{2},p_{3})\,{\cal R}_{B_{cll'}}(p_{1},p_{2},p_{3};p).\label{eq:fbc}
\end{align}
Here, $f_{c\bar{l}}(p_{1},p_{2})$ is the joint momentum distribution
for charm ($c$) quark and light anti-quark ($\bar{l}$). ${\cal R}_{M_{c\bar{l}}}(p_{1},p_{2};p)$
is the combination function that is the probability density for the
given $c\bar{l}$ with momenta $p_{1}$, $p_{2}$ combining into a
meson $M_{c\bar{l}}$ with momentum $p$. It is similar for the baryon. 

We take independent distributions for quarks of different flavors
by neglecting correlations, 
\begin{align}
f_{c\bar{l}}(p_{1},p_{2}) & =f_{c}(p_{1})f_{\bar{l}}(p_{2}),\label{eq:factcl}\\
f_{cll'}(p_{1},p_{2},p_{3}) & =f_{c}(p_{1})f_{l}(p_{2})f_{l'}(p_{3}).\label{eq:factcll}
\end{align}
We suppose the combination takes place mainly for quark and/or antiquark
that takes a given fraction of momentum of the hadron so that the
combination function is the product of Dirac delta functions
\begin{align}
{\cal R}_{M_{c\bar{l}}}(p_{1},p_{2};p) & =\kappa_{M_{c\bar{l}}}\prod_{i=1}^{2}\delta(p_{i}-x_{i}p),\label{eq:rmc}\\
{\cal R}_{B_{cll'}}(p_{1},p_{2},p_{3};p) & =\kappa_{B_{cll'}}\prod_{i=1}^{3}\delta(p_{i}-x_{i}p),\label{eq:rbc}
\end{align}
where $\kappa_{M_{c\bar{l}}}$ and $\kappa_{B_{cll'}}$ are constants
which are independent of the momentum but dependent on other ingredients
such as the quark number so that all charm quarks can be correctly
exhausted (after further including multi-charm hadrons). 

Following our works \cite{Song:2017gcz,Gou:2017foe} for light-flavor
hadrons in $pp$ and $p$-Pb collisions at LHC energies, we adopt
the co-moving approximation in combination, i.e., the charm quark
combines with light quark(s) of the same velocity to form the charm
hadron. Since the equal velocity implies $p_{i}=\gamma vm_{i}\propto m_{i}$,
the momentum fraction is
\begin{equation}
x_{i}=m_{i}/\sum_{j}m_{j},
\end{equation}
where quark masses are taken to be $m_{u}=m_{d}=0.33$ GeV, $m_{s}=0.5$
GeV, and $m_{c}=1.5$ GeV, the constituent masses in the quark model.\textbf{
}Substituting Eqs. (\ref{eq:factcl}-\ref{eq:factcll}) and (\ref{eq:rmc}-\ref{eq:rbc})
into Eqs. (\ref{eq:fmc}-\ref{eq:fbc}), we obtain the distributions
of single-charm hadrons
\begin{align}
f_{M_{c\bar{l}}}(p) & =\kappa_{M_{c\bar{l}}}f_{c}(x_{1}p)f_{\bar{l}}(x_{2}p),\label{eq:fmz-1}\\
f_{B_{cll'}}(p) & =\kappa_{B_{cll'}}f_{c}(x_{1}p)f_{l}(x_{2}p)f_{l'}(x_{3}p).\label{eq:fbz-1}
\end{align}

We rewrite the distribution functions of charm hadrons, 
\begin{align}
f_{M_{c\bar{l}}}\left(p\right) & =N_{M_{c\bar{l}}}\,f_{M_{c\bar{l}}}^{\left(n\right)}\left(p\right),\label{eq:fmfinal}\\
f_{B_{cll'}}\left(p\right) & =N_{B_{cll'}}\,f_{B_{cll'}}^{\left(n\right)}\left(p_{}\right),\label{eq:fbfinal}
\end{align}
where $f_{M_{c\bar{l}}}^{\left(n\right)}\left(p\right)$ is the normalized
distribution function with $\int dp\,f_{M_{c\bar{l}}}^{\left(n\right)}\left(p\right)=1$.
$N_{M_{c\bar{l}}}$ is momentum-integrated yield 
\begin{align}
N_{M_{c\bar{l}}} & =N_{c}N_{\bar{l}}\,\frac{\kappa_{M_{c\bar{l}}}}{A_{M_{c\bar{l}}}}=N_{c}N_{\bar{l}}\,\mathcal{R}_{c\bar{l}\rightarrow M_{c\bar{l}}},\\
N_{B_{cll'}} & =N_{c}N_{l}N_{l'}\frac{\kappa_{B_{cll'}}}{A_{B_{cll'}}}=N_{c}N_{l}N_{l'}\mathcal{R}_{cll'\rightarrow B_{cll'}},
\end{align}
 where $A_{M_{c\bar{l}}}^{-1}=\int dp\,f_{c}^{\left(n\right)}\left(x_{1}p\right)f_{\bar{l}}^{\left(n\right)}\left(x_{2}p\right)$
and $A_{B_{cll'}}^{-1}=\int{\rm d}p\prod_{i=1}^{3}f_{q_{i}}^{\left(n\right)}\left(x_{i}p\right)$
with the normalized charm and light quark distribution $\int dp\,f_{c,l}^{\left(n\right)}\left(p\right)=1$.
We see that $\mathcal{R}_{c\bar{l}\rightarrow M_{c\bar{l}}}\equiv\kappa_{M_{c\bar{l}}}/A_{M_{c\bar{l}}}$
is nothing but the momentum-integrated combination probability of
$c\bar{l}\rightarrow M_{c\bar{l}}$. It is similar for $\mathcal{R}_{cll'\rightarrow B_{cll'}}\equiv\kappa_{B_{cll'}}/A_{B_{cll'}}$. 

$\mathcal{R}_{c\bar{l}\rightarrow M_{c\bar{l}}}$ and $\mathcal{R}_{cll'\rightarrow B_{cll'}}$
are parameterized. We use $N_{M_{c}}$ to denote the total number
of all single-charm mesons. $N_{c\bar{q}}=N_{c}\left(N_{\bar{u}}+N_{\bar{d}}+N_{\bar{s}}\right)$
is the possible number of all charm-light pairs. $N_{M_{c}}/N_{c\bar{q}}$
gives the flavor-averaged probability of a $c\bar{l}$ forming a charm
meson. The average number of $M_{c\bar{l}}$ is $N_{c}N_{\bar{l}}\times\left(N_{M_{c}}/N_{c\bar{q}}\right)=P_{\bar{l}}N_{M_{c}}$
where $P_{\bar{l}}\equiv N_{\bar{l}}/N_{\bar{q}}$ denotes the probability
of an antiquark with the flavor $\bar{l}$. For a given $c\bar{l}$
combination, it can form different $J^{P}$ states, and we use $C_{M_{i,c\bar{l}}}$
to denote the probability of forming the particular spin state $i$,
and finally obtain the yield of charm meson $M_{i,c\bar{l}}$ 
\begin{equation}
N_{M_{i,c\bar{l}}}=C_{M_{i,c\bar{l}}}P_{\bar{l}}N_{M_{c}}.\label{eq:nmi}
\end{equation}
In this paper we consider only the pseudo-scalar mesons $J^{P}=0^{-}$($D^{+}$,
$D^{0}$ and $D_{s}^{+}$ ) and vector mesons $J^{P}=1^{-}$ ($D^{*+}$,
$D^{*0}$ and $D_{s}^{*+}$ ) in the ground state. We introduce a
parameter $R_{V/P}$ to denote the relative ratio of vector meson
to pseudo-scalar meson of the same quark flavors, and have 
\begin{equation}
C_{M_{i,c\bar{l}}}=\begin{cases}
\frac{1}{1+R_{V/P}} & \text{for}\,J^{P}=0^{-}\,\text{mesons}\\
\frac{R_{V/P}}{1+R_{V/P}} & \text{for}\,J^{P}=1^{-}\,\text{mesons}.
\end{cases}
\end{equation}
We take $R_{V/P}=1.5$, the thermal weight value used in \cite{Rapp:2003wn,Oh:2009zj,Li:2017zuj}.

In baryon sector, we have 
\begin{equation}
N_{B_{i,cll'}}=C_{B_{i,cll'}}N_{iter,ll'}P_{l}P_{l'}N_{B_{c}},\label{eq:nbi}
\end{equation}
 where $N_{B_{c}}$ is the total number of all single-charm baryons,
$N_{iter,ll'}P_{l}P_{l'}$ selects the specific light flavor $ll'$,
and $C_{B_{i,cll'}}$ selects the particular spin state. Here $P_{l}=N_{l}/N_{q}=N_{l}/\left(N_{u}+N_{d}+N_{s}\right)$
denotes the probability of a quark with the flavor $l$. $N_{iter,ll'}$
is the iteration number of $ll'$ pair and is taken to be 1 for $l=l'$
and 2 for $l\neq l'$. We consider the production of triplet ($\Lambda_{c}^{+},\,\Xi_{c}^{+},\,\Xi_{c}^{0}$)
with $J^{P}=\left(1/2\right)^{+}$, sextet $\left(\Sigma_{c}^{0},\,\Sigma_{c}^{+},\,\Sigma_{c}^{++},\,\Xi_{c}^{'0},\,\Xi_{c}^{'+},\,\Omega_{c}^{0}\right)$
with $J^{P}=\left(1/2\right)^{+}$, and sextet $\left(\Sigma_{c}^{*0},\,\Sigma_{c}^{*+},\,\Sigma_{c}^{*++},\,\Xi_{c}^{*0},\,\Xi_{c}^{*+},\,\Omega_{c}^{*0}\right)$
with $J^{P}=\left(3/2\right)^{+}$, respectively, in the ground state.
We introduce a parameter $R_{S1/T}$ to denote the relative ratio
of $J^{P}=\left(1/2\right)^{+}$ sextet baryons to $J^{P}=\left(1/2\right)^{+}$
triplet baryons of the same quark flavors, and a parameter $R_{S3/S1}$
to denote that of $J^{P}=\left(3/2\right)^{+}$ sextet baryons to
$J^{P}=\left(1/2\right)^{+}$ sextet baryons of the same quark flavors.
We also take the the effective thermal weighs as a guideline and take
$R_{S1/T}=0.5$ and $R_{S3/S1}=1.5$, respectively. For $ll'=uu,dd,ss$,
\begin{equation}
C_{B_{i,cll'}}=\begin{cases}
\frac{1}{1+R_{S3/S1}} & \text{for}\,\Sigma_{c}^{++},\Sigma_{c}^{0},\,\Omega_{c}^{0}\\
\frac{R_{S3/S1}}{1+R_{S3/S1}} & \text{for}\,\Sigma_{c}^{*++},\Sigma_{c}^{*0},\,\Omega_{c}^{*0}.
\end{cases}
\end{equation}
For $ll'=ud,us,ds$, 
\begin{equation}
C_{B_{i,cll'}}=\begin{cases}
\frac{1}{1+R_{S1/T}\left(1+R_{S3/S1}\right)} & \text{for}\,\Lambda_{c}^{+},\,\Xi_{c}^{0},\,\Xi_{c}^{+}\\
\frac{R_{S1/T}}{1+R_{S1/T}\left(1+R_{S3/S1}\right)} & \text{for}\,\Sigma_{c}^{+},\,\Xi_{c}^{'0},\,\Xi_{c}^{'+}\\
\frac{R_{S1/T}R_{S3/S1}}{1+R_{S1/T}\left(1+R_{S3/S1}\right)} & \text{for}\,\Sigma_{c}^{*+},\,\Xi_{c}^{*0},\,\Xi_{c}^{*+}.
\end{cases}
\end{equation}
 We note that yields and momentum spectra of final state $\Lambda_{c}^{+}$,
$\Xi_{c}^{0}$ and $\Omega_{c}^{0}$ after taking decay contribution
into account are not sensitive to parameters $R_{S1/T}$ and $R_{S3/S1}$. 

Considering the single-charm mesons and baryons consume most of charm
quarks produced in collisions, we have the following approximated
normalization to single-charm hadrons 
\begin{equation}
N_{M_{c}}+N_{B_{c}}\approx N_{c}.\label{eq:nc}
\end{equation}
Here we treat the ratio $R_{B/M}^{\left(c\right)}\equiv N_{B_{c}}/N_{M_{c}}$
as a parameter of the model, which characterizes the relative production
of single-charm baryons to single-charm mesons. We take $R_{B/M}^{\left(c\right)}=0.425$,
the value in previous work \cite{Li:2017zuj}. 

\section{Results and discussions }

We apply the above formulas in QCM to the one-dimensional $p_{T}$
space and calculate the $p_{T}$ spectra of single-charm hadrons at
mid-rapidity in $pp$ collisions at $\sqrt{s}=7$ TeV. The $p_{T}$
distributions of quarks at hadronization are inputs of the model.
We have obtained the $p_{T}$ spectra of light-flavor quarks in previous
work \cite{Gou:2017foe}. The averaged quark number in the rapidity
interval $|y|<0.5$ is 2.5 for $u$ quark and 0.8 for $s$ quark,
respectively. The normalized distributions $f_{u}^{\left(n\right)}\left(p_{T}\right)$
and $f_{s}^{\left(n\right)}\left(p_{T}\right)$ are shown in Fig.1(a).
The charge conjugation symmetry between quark and antiquark and the
iso-spin symmetry between up and down quarks are applied in calculations. 

\begin{figure}[tbh]
\includegraphics[scale=0.4]{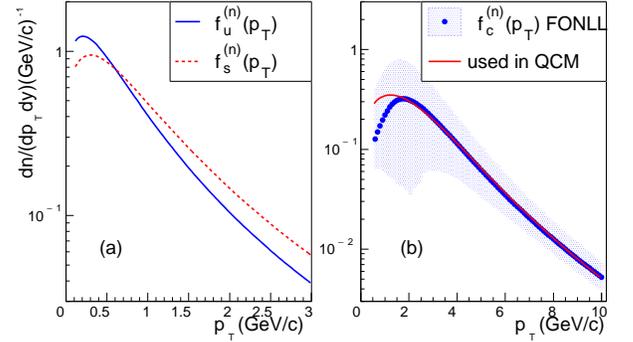}\caption{(a) The normalized $p_{T}$ spectra of light quarks at mid-rapidity
in $pp$ collisions at $\sqrt{s}=7$ TeV;(b) that of charm quarks.
The shadow area shows the scale uncertainties in FONLL calculation.}
\label{fig1}
\end{figure}

In Fig. \ref{fig1}(b), we show the normalized distribution of charm
quarks, which is obtained from the online calculation of Fixed-Order
Next-to-Leading-Logarithmic (FONLL) \footnote{FONLL Heavy Quark Production, http://www.lpthe.jussieu.fr/\textasciitilde{}cacciari/fonll/fonllform.html}.
The points are center values and the shadow area shows the scale uncertainties,
see Refs. \cite{Cacciari:1998it,Cacciari:2012ny} for details. The
uncertainty due to parton distribution functions (PDFs) is not included.
Because of the large theoretical uncertainty, in particular, at low
$p_{T}$, we only take the FONLL calculation as a guideline. The practically
used $p_{T}$ spectrum of charm quarks is reversely extracted from
the data of $D^{*+}$ meson \cite{Abelev:2014hha,Adam:2016ich} in
QCM and is shown as the thick solid line in Fig. \ref{fig1}(b). The
cross section of charm quarks in $|y|<0.5$ interval is 1.2 mb. The
extracted spectrum is found to be very close to the center values
of FONLL calculation for $p_{T}\gtrsim1.5$ GeV/$c$ and be higher
than the latter to a certain extent for $p_{T}<1.5$ GeV/$c$ but
be still within the theoretical uncertainties.

\begin{figure}[tbh]
\includegraphics[scale=0.4]{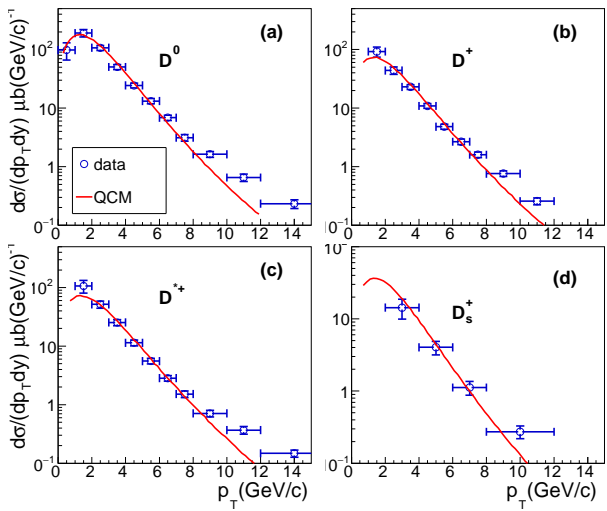}\caption{Differential cross sections of $D$ mesons at mid-rapidity as the
function of $p_{T}$ in $pp$ collisions at $\sqrt{s}=7$ TeV. Symbols
are experimental data \cite{Acharya:2017jgo} and lines are results
of QCM. }
\label{fig2}
\end{figure}

In Fig. \ref{fig2}, we show results of differential cross sections
of $D$ mesons at mid-rapidity as the function of $p_{T}$ in $pp$
collisions at $\sqrt{s}=7$ TeV, and compare them with experimental
data \cite{Acharya:2017jgo}. We see that QCM well describes the data
of $D$ mesons for $p_{T}\lesssim7$ GeV/$c$ but under-predicts the
data for larger $p_{T}$. This is reasonable. In the equal-velocity
combination of charm quarks and light quarks, a charm quark with $p_{T,c}\lesssim6$
GeV/$c$ will combine a light antiquark of $p_{T,\bar{l}}\lesssim1.5$
GeV/$c$. Because most of light quarks, see Fig. \ref{fig1}(a), are
of such low $p_{T}$, they provide the sufficient partners (or chance)
for the hadronization of charm quarks. For a charm quark of $p_{T,c}\gtrsim6$
GeV/$c$, the combining light antiquark should have $p_{T,\bar{l}}\gtrsim1.5$
GeV/$c$ whose number is very small and is exponentially dropped.
In this case, those light antiquarks may be not enough to provide
the sufficient chance for the combination hadronization of charm quarks
of $p_{T,c}\gtrsim6$ GeV/$c$, and therefore the combination may
be not the dominated channel and the fragmentation will take over.

\begin{figure}[tbh]
\includegraphics[scale=0.4]{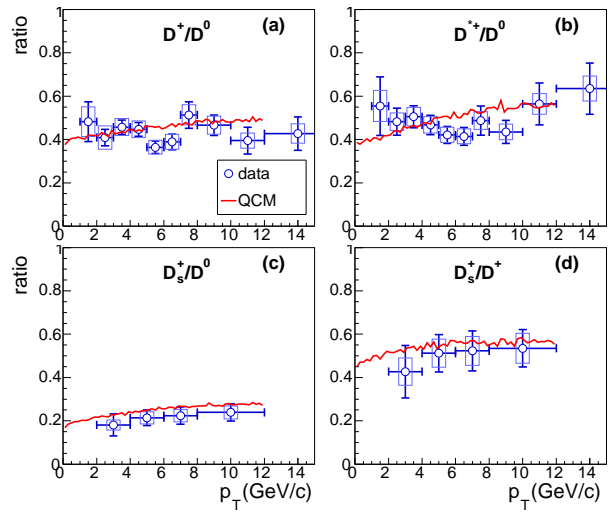}\caption{ Ratios of different $D$ mesons as the function of $p_{T}$ in $pp$
collisions at $\sqrt{s}=7$ TeV. Symbols are experimental data \cite{Acharya:2017jgo}
and lines are results of QCM. }
\label{fig3}
\end{figure}

In Fig. \ref{fig3}, we show results for the ratios of different $D$
mesons as the function of $p_{T}$ in $pp$ collisions at $\sqrt{s}=7$
TeV, and compare them with experimental data \cite{Acharya:2017jgo}.
We see that, within experimental uncertainties, the model results
are in agreement with the data. For the magnitudes of these four ratios,
we can give a simple explanation from the yield (corresponding to
differential cross-section) ratios of $D$ mesons. Using Eq. (\ref{eq:nmi})
and taking strong and electromagnetic decay contribution into account
where the data of decay branch ratios are taken from PDG\cite{PDG:2016xqp},
we have
\begin{align}
\frac{D^{+}}{D^{0}} & =\frac{1+0.323R_{V/P}}{1+1.677R_{V/P}}\approx0.42,\label{eq:ratio_Dp_D0}\\
\frac{D^{*+}}{D^{0}} & =\frac{R_{V/P}}{1+1.677R_{V/P}}\approx0.43,\label{eq:ratio_Dstar_D0-1}\\
\frac{D_{s}^{+}}{D^{0}}= & \frac{1+R_{V/P}}{1+1.677R_{V/P}}\lambda_{s}\approx0.23,\label{eq:ratio_Ds_D0}\\
\frac{D_{s}^{+}}{D^{+}} & =\frac{1+R_{V/P}}{1+0.323R_{V/P}}\lambda_{s}\approx0.54\label{eq:ratio_Ds_Dplus}
\end{align}
with $\lambda_{s}=N_{s}/N_{u}=0.32$ and $R_{V/P}=1.5$. 

Theoretical pQCD calculations with fragmentation functions were compared
with experimental data of $D$ mesons in Ref. \cite{Acharya:2017jgo}.
It is shown that pQCD calculations in large $p_{T}$ range have small
theoretical uncertainties and often well explain the data. However,
pQCD calculations in small $p_{T}$ range have quite large theoretical
uncertainties and the comparison with data is not conclusive.\emph{
}In contrast with those pQCD calculations with fragmentation functions,
our results suggest a different mechanism for the charm quark hadronization
at low $p_{T}$. 

\begin{figure}[tbh]
\includegraphics[scale=0.4]{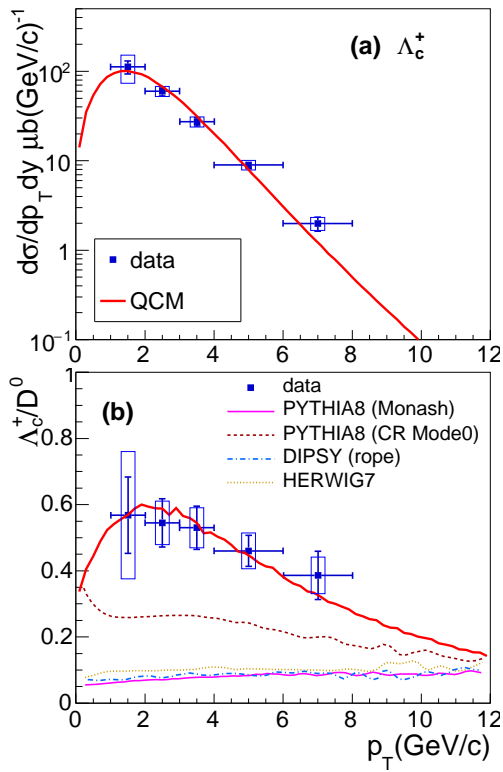}\caption{ Differential cross-section of $\Lambda_{c}^{+}$ at mid-rapidity
(a) and the ratio to $D^{0}$ (b) as the function of $p_{T}$ in $pp$
collisions at $\sqrt{s}=7$ TeV. Symbols are experimental data \cite{Acharya:2017kfy}
and the thick solid lines are results of QCM. Results of other models
or event generators in panel (b) are taken from \cite{Acharya:2017kfy}. }
\label{fig4}
\end{figure}

The production of baryons is more sensitive to the hadronization mechanism.
In Fig. \ref{fig4}, we show results of the $p_{T}$ spectrum of baryon
$\Lambda_{c}^{+}$ and the ratio to $D^{0}$ meson, and compare them
with the experimental data \cite{Acharya:2017kfy}. We see that, similar
to $D$ mesons, our results of $\Lambda_{c}^{+}$ spectrum and ratio
$\Lambda_{c}^{+}/D^{0}$ are in good agreement with the data for $p_{T}\lesssim7$
GeV/$c$. The predictions of other models or event generators \cite{Christiansen:2015yqa,Bierlich:2015rha,Bahr:2008pv,Sjostrand:2007gs}
which adopt string or cluster fragmentation mechanism for hadronization
are also shown in Fig. \ref{fig4}(b). PYTHIA8 Monash tune, DIPSY
with rope parameter, and HERWIG7 predict a small $\Lambda_{c}^{+}/D^{0}$
ratio of 0.1 and almost constant ratio at different $p_{T}$. Considering
the effect of string formation beyond leading color approximation,
PYTHIA8 (CR Mode0) increases the ratio to a certain extent and gives
the decrease tendency with $p_{T}$. 

\begin{figure}[tbh]
\includegraphics[scale=0.4]{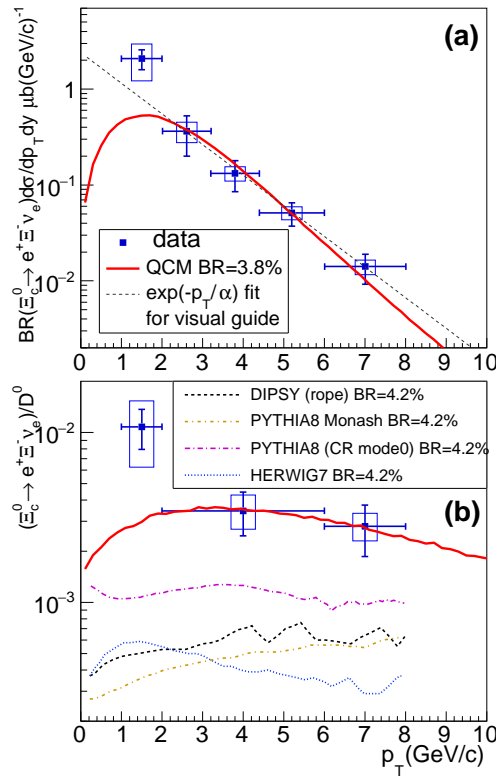}\caption{Differential cross-section of $\Xi_{c}^{0}$ at mid-rapidity multiplied
by the branch ratio into $e^{+}\Xi^{-}\nu_{e}$ (a) and the ratio
to $D^{0}$ (b) as the function of $p_{T}$ in $pp$ collisions at
$\sqrt{s}=7$ TeV. Symbols are experimental data \cite{Acharya:2017lwf}
and the thick solid lines are results of QCM. Results of other models
or event generators in panel (b) are taken from \cite{Acharya:2017lwf}. }
\label{fig5}
\end{figure}

In Fig. \ref{fig5}, we show results of the differential cross-section
of $\Xi_{c}^{0}$ at mid-rapidity multiplied by the branch ratio into
$e^{+}\Xi^{-}\nu_{e}$ (a) and the relative ratio to $D^{0}$ (b)
as the function of $p_{T}$ in $pp$ collisions at $\sqrt{s}=7$ TeV.
Because of the lack of the absolute branch ratio into $e^{+}\Xi^{-}\nu_{e}$,
our result of the spectrum of $\Xi_{c}^{0}$ is multiplied by a branch
ratio 3.8\%, which is within the current range of theoretical calculations
(0.83\%-4.2\%). We see that the model result, the thick solid line
in Fig. \ref{fig5}(a), can well describe the data of $\Xi_{c}^{0}$
for $2\lesssim p_{T}\lesssim7$ GeV/$c$ but significantly underestimates
the first data point at $p_{T}=1.5$ GeV/$c$. However, the first
data point, to our knowledge, is somewhat puzzlingly high if we note
that the studied differential cross-section is $d\sigma/dp_{T}dy$.
The data of $\Lambda_{c}^{+}$ in Fig. \ref{fig4}(a) and $D^{0}$
in Fig. \ref{fig2}(a) suggest that the differential cross-section
tends to increase slowly with the decreasing $p_{T}$ for small $p_{T}\lesssim2$
GeV/$c$ and will saturate and decrease as $p_{T}\to0$. The data
of light-flavor hadrons for $dN/dp_{T}dy$, e.g. $K\left(892\right)^{*}$,
show this behavior more clearly \cite{Abelev:2012hy}. As a naive
illustration, we see the first data point of $\Xi_{c}^{0}$ at $p_{T}=1.5$
GeV/$c$ is more than twice the exponential extrapolation from data
points of larger $p_{T}$, the thin dashed line, which is not the
case for the data of $D$ mesons and $\Lambda_{c}^{+}$. The QCM result
of the ratio $\Xi_{c}^{0}/D^{0}$ is shown in Fig. \ref{fig5}(b).
We see that the two data points within $2\lesssim p_{T}\lesssim7$
GeV/$c$ can be well described by QCM and the first data point at
$p_{T}=1.5$ GeV/$c$ is much higher than the QCM result. 

String/cluster fragmentation usually under-predicts the production
of $\Xi_{c}^{0}$. Here, we show predictions of several models or
event generators which adopt string/cluster fragmentation at hadronization.
They are taken from Ref. \cite{Acharya:2017lwf} and are shown as
different kinds of thin lines in Fig. \ref{fig5}(b). The decay branch
$\Xi_{c}^{0}\to e^{+}\Xi^{-}\nu_{e}$ is taken to be 4.2\%, and therefore
these predictions correspond to the up limits. HERWIG7 which adopts
the cluster fragmentation predicts the decreasing ratio for $p_{T}\gtrsim1$
GeV/$c$ but is lower than the data about an order of the magnitude.
PYTHIA8 (Monash tune) and DIPSY with rope parameter which adopt sting
fragmentation predict the increasing ratio as the function of $p_{T}$
and are significantly lower than the data. PYTHIA8 (CR mode0) which
takes the color re-connection into account by considering the string
formation beyond leading color approximation will increase the prediction
of ratio to a large extent but the prediction is still only one third
of the data. 

\section{Summary and discussion}

We have shown the experimental data of $p_{T}$ spectra of single-charm
hadrons $D^{0,+}$, $D^{*+}$ $D_{s}^{+}$, $\Lambda_{c}^{+}$ and
$\Xi_{c}^{0}$ at mid-rapidity in the low $p_{T}$ range ($2\lesssim p_{T}\lesssim7$
GeV/$c$) in $pp$ collisions at $\sqrt{s}=7$ TeV can be well understood
by the equal-velocity combination of perturbatively-created charm
quarks and the light-flavor constituent quarks and antiquarks. We
emphasize the following aspects to address the physical importance
of our results. (1) The property, i.e., the $p_{T}$ distributions,
of light-flavor constituent quarks and antiquarks at hadronization
are obtained from the data of $p_{T}$ spectra of light-flavor hadrons
in work \cite{Gou:2017foe} where it is found that equal-velocity
combination of light-flavor quarks can reasonably describe the data
of light flavor hadrons in the low $p_{T}$ range. The existence of
the underlying source of light-flavor quarks is a new property of
small parton system, maybe related to the creation of the deconfined
parton system in $pp$ collisions at LHC energies. (2) The good performance
for the combination of charm quarks and those constituent quarks and
antiquarks in comparison with the data suggests a new scenario of
the low $p_{T}$ charm quark hadronization in the presence of the
underlying light quark source in $pp$ collisions at LHC energies,
in contrast to the usually adopted fragmentation mechanism. (3) Most
of light quarks combine into light-flavor hadrons that reproduces
the data of light-flavor hadrons. A small fraction of light quarks
combine with charm quarks which also well explains the data of single-charm
hadrons in low $p_{T}$ range. This provides a possible universal
picture for the production of low $p_{T}$ hadrons in $pp$ collisions
at LHC energies. 

Several discussions on the limitation of our model and results are
necessary. The present work only focuses on the characteristic of
hadron production in (transverse) momentum space. It is still unclear
that what kind of the spatial property for the (light-flavor dominated)
small parton system leads to the effective combination of charm quarks
and those light-flavor quarks. In particular, in the light-flavor
sector we adopt the concept of the constituent quarks. What kind of
the spatial property for the small parton system is responsible for
the exhibition of the light-flavor constituent quarks degrees of freedom?
Is it related to the possible de-confinement in the small system of
$pp$ collisions at LHC energies? These interesting and important
questions are deserved to study in future. 
\begin{acknowledgments}
This work is supported by the National Natural Science Foundation
of China under Grant Nos. 11575100.\\
\end{acknowledgments}

\bibliographystyle{apsrev4-1}
\bibliography{ref}

\end{document}